# Transforming disaster risk reduction with AI and big data: Legal and interdisciplinary perspectives

Kwok P Chun[1], Thanti Octavianti[2], Nilay Dogulu[3], Hristos Tyralis[4], Georgia Papacharalampous[5], Ryan Rowberry[6], Pingyu Fan[7], Mark Everard[8], Maria Francesch-Huidobro[9], Wellington Migliari[10], David M. Hannah[11], John Travis Marshall[12], Rafael Tolosana Calasanz[13], Chad Staddon[14], Ida Ansharyani[15], Bastien Dieppois[16], Todd R Lewis[17], Juli Ponce[18], Silvia Ibrean[19], Tiago Miguel Ferreira[20], Chinkie Peliño-Golle[21], Ye Mu[22], Manuel Delgado[23], Elizabeth Silvestre Espinoza[24], Martin Keulertz[25], Deepak Gopinath[26], Cheng Li[27]

[1] School of Architecture and Environment, University of the West of England, UK, kwok.chun@uwe.ac.uk, https://orcid.org/0000-0001-9873-6240
[2] School of Architecture and Environment, University of the West of England, UK, thanti.octavianti@uwe.ac.uk, https://orcid.org/0000-0002-9921-0539
[3] World Meteorological Organization, Genève, Switzerland nilay.dogulu@gmail.com, https://orcid.org/0000-0003-4229-2788
[4] Construction Agency, Hellenic Air Force, Mesogion Avenue 227-231, 15561, Cholargos, Greece, montchrister@gmail.com, https://orcid.org/0000-0002-8932-4997
[5] Department of Topography, School of Rural, Surveying and Geoinformatics Engineering, National Technical University of Athens, Greece, papacharalampous.georgia@gmail.com, https://orcid.org/0000-0001-5446-954X
[6] Georgia State University College of Law, Atlanta, Georgia, USA, rrowberry@gsu.edu
[7] Department of Geography, Hong Kong Baptist University, Hong Kong, 18482392@life.hkbu.edu.hk
[8] School of Architecture and Environment, University of the West of England, UK, mark.everard@uwe.ac.uk, https://orcid.org/0000-0002-0251-8482
[9] Department of Geography, The University of Hong Kong, fmariade@hku.hk
[10] Law School, University of Barcelona, Catalonia, Spain, wemigliari@ub.edu, https://orcid.org/0000-0002-1073-1312
[11] School of Geography, Earth & Environmental Sciences, University of Birmingham, UK, d.m.hannah@bham.ac.uk, https://orcid.org/0000-0003-1714-1240
[12] Georgia State University College of Law, USA, jmarshall32@gsu.edu, https://orcid.org/0000-0003-0016-2125
[13] Department of Computer Science and Systems Engineering, Universidad de Zaragoza, Spain, rafaelt@unizar.es
[14] School of Architecture and Environment, University of the West of England, UK, Chad.Staddon@uwe.ac.uk, https://orcid.org/0000-0002-2063-8525
[15] Department of Agriculture, Universitas Samawa, Indonesia, idaansharyani@gmail.com
[16] Centre for Agroecology, Water and Resilience (CAWR), Coventry University, Coventry, UK, ab9482@coventry.ac.uk
[17] University of the West of England, todd.lewis@uwe.ac.uk, https://orcid.org/0000-0001-5433-8777
[18] Law School, University of Barcelona, Catalonia, Spain, jponce@ub.edu, https://orcid.org/0000-0002-1977-5063

[19] UN Volunteer, United Nations Volunteers, Hermann-Ehlers-Strasse 10, Bonn, Germany, silvia.ibrean@yahoo.com, https://orcid.org/0009-0004-3553-0041
[20] College of Arts, Technology and Environment, School of Engineering, University of the West of England, Tiago.Ferreira@uwe.ac.uk, https://orcid.org/0000-0001-6454-7927
[21] EcoWaste Coalition, Philippines, chinkiep@gmail.com
[22] Department of Geography, University of California, Santa Barbara, USA, ymu@ucsb.edu
[23] University of the West of England, Faculty of Business and Law, manuel.daviladelgado@uwe.ac.uk   manuel.daviladelgado@uwe.ac.uk
[24] Inclima, elizabeth.silvestre@inclima.cc, https://orcid.org/0000-0003-1149-8804
[25] School of Architecture and Environment, University of the West of England, martin.keulertz@uwe.ac.uk
[26] School of Architecture and Environment, University of the West of England, Deepak.Gopinath@uwe.ac.uk
[27] Department of Ecology, Yangzhou University, China, licheng@yzu.edu.cn

**Abstract:** Managing complex disaster risks requires interdisciplinary efforts. Breaking down silos between law, social sciences, and natural sciences is critical for all processes of disaster risk reduction. This enables adaptive systems for the rapid evolution of AI technology, which has significantly impacted the intersection of law and natural environments. Exploring how AI influences legal frameworks and environmental management, while also examining how legal and environmental considerations can confine AI within the socioeconomic domain, is essential.

From a co-production review perspective, drawing on insights from lawyers, social scientists, and environmental scientists, principles for responsible data mining are proposed based on safety, transparency, fairness, accountability, and contestability. This discussion offers a blueprint for interdisciplinary collaboration to create adaptive law systems based on AI integration of knowledge from environmental and social sciences. Discrepancies in the use of language between environmental scientists and decision-makers in terms of usefulness and accuracy hamper how AI can be used based on the principles of legal considerations for a safe, trustworthy, and contestable disaster management framework.

When social networks are useful for mitigating disaster risks based on AI, the legal implications related to privacy and liability of the outcomes of disaster management must be considered. Fair and accountable principles emphasise environmental considerations and foster socioeconomic discussions related to public engagement. AI also has an important role to play in education, bringing together the next generations of law, social sciences, and natural sciences to work on interdisciplinary solutions in harmony.

Although emerging AI approaches can be powerful tools for disaster management, they must be implemented with ethical considerations and safeguards to address concerns about bias, transparency, and privacy. The responsible execution of AI approaches, based on the dynamic interplay between AI, law, and environmental risk, promotes sustainable and equitable practices in data mining.

## 1. INTRODUCTION

Since the 2010s, governments across the world have developed and applied various forms of Artificial Intelligence (AI) to address intersectoral priorities of legal and environmental systems. Today, the emerging frontier has morphed from implementation to regulation of the rapid roll-out of AI (Smuha, 2019). Emerging data-driven technologies are rapidly transforming the ways in which we live and work. Disaster risk management is no exception. With more frequent and severe environmental disasters, AI is increasingly used as a tool for the adaptive data mining management of disaster risks (e.g. Yu et al., 2018; Chen et al., 2019; Imran et al., 2020; Fan & Chun, 2022; Alizadeh et al., 2022; Chauhan et al 2024) from developing early warning systems, providing real-time situational awareness during disasters, to assisting aid allocation post-disasters. However, AI also poses significant risks. For example, since AI systems rely on historical data to facilitate decision-making, incomplete or bias data could result in biased outcomes. This is particularly important for disaster management given that AI-assisted decisions could potentially have high-impact outcomes on people's lives and the environment they live in (Gevaert et al., 2021). Regulators are, therefore, urged to look beyond the benefits of AI and data mining, putting in place appropriate measures to ensure its responsible and accountable use. However, AI models have applications beyond public sector control, also informing corporate decision-making (particularly in the light of recent uptake of Taskforce on Nature-related Financial Disclosures (TNFD) and other voluntary pro-environmental measures) as well as surveillance by NGOs and other sector of civil society.

In the field of disaster management, AI methods can be classified between two extremes: supervised learning and unsupervised learning (Guikema, 2020). Under supervised learning, a correct or 'desired' answer is provided to the algorithm, providing a reference point to check through large disaster datasets. An example would be in recording a higher-than-normal day temperatures over a week and then linking it to a 'correct answer', i.e., a prediction of increase in hospital admissions prior to the summer heatwave. In unsupervised learning methods, no correct or 'desired' representations are provided to the algorithm to seek to develop in-depth understanding of the interplay between disaster management variables. Beyond supervised and unsupervised methods are a class of methods that can be grouped under 'deep learning', where methods such as 'recursive neural networks', 'reinforcement learning' in resource management, etc. have been used, which require complex sequential tasks, large amounts of training data and time for disaster risk reduction (Sun et al., 2020). In general, learning methods underpinning the application of AI are more likely to generate useful information, when large amount of data is generated repeatedly in similar locations or scenarios. However, when enough volume of data is not available, the AI model prediction can be inaccurate due to inadequate training and insufficient learning from data provided for disaster management (Guikema, 2020).

Although AI has been deployed to address the global agenda for sustainable development (Vinuesa et al., 2020), the practice of AI and data mining has only started receiving attention from a regulatory perspective in support of its responsible (ethical) disaster management (e.g. Siau & Wang, 2020; Deltares, WB & GFDRR, 2021). Regulators at national, regional, international and supranational levels have started assessing the necessity of revising existing regulations or developing novel regulatory approaches to mitigate AI risks. As an evolving and complex technology, AI poses multiple unique regulatory challenges. These include transboundary use of AI making it difficult to enforce regulations; and constrictive requirements that potentially curb innovations (Coeckelbergh, 2019). To balance the protective and enabling roles of regulation, various countries are considering following a principle-based approach (focusing on the outcomes regardless of the process or means), a rule-based approach (focusing on the process regardless of the outcomes), or a mix of both approaches (Frantz & Instefjord, 2018). For example, the European Commission (EC) is proposing the first-ever legal framework on AI by classifying systems into four levels of risk, ranging from minimal or no risk to unacceptable risk (EC, 2023). China is also in this race with its emerging strategies and detailed regulation to govern AI (Trustible, 2023). The United Kingdom (UK) is taking a pro-innovation approach by using five principles to guide the responsible development and ethical use of AI and data mining in all policy sectors (DSIT, 2023). The US is in the process of reviewing its current approach to AI regulation and the development of a risk-based framework for AI (Seamans, 2023). Many of these policy frameworks have been shaped by IEEE's (2019) report, 'Ethically Aligned Design', which brings to the fore ethical considerations while developing and deploying 'Autonomous and Intelligence Systems' including AI tools.

All these initiatives emphasise the need for adaptive law and policy systems. However, there is a missing conversation around how disciplines with a stake on AI and disaster management can (and should) shape the momentum towards this goal. Taking the lens of adaptive intersection between legal and environmental systems for disaster management as a case, we present our shared insights that resulted from a series of co-production workshops and focus-groups.[1] We analyse AI regulatory challenges and desirability in the disaster risk reduction sector based on the five principles of ethical AI (DSIT, 2023), introduced in Section 2. Section 3 will then address the wider significance of these principles in disaster management across three areas: building adaptive systems for legal and physical environments; encouraging public engagement and law; and forging interdisciplinary training in disaster management fields.

## 2. PRINCIPLES OF ETHICAL AI USE FOR DISASTER MANAGEMENT

In this section, we describe the principles of ethical AI with a focus on disaster management in the context of environmental hazards (Figure 1). To establish a common ground for a joint conversation on AI regulatory challenges and desirability in the disaster management, we followed the five principles of ethical AI introduced by the UK (DSIT, 2023) comprising: 1) safety, security and robustness; 2) transparency and explainability; 3) fairness; 4) accountability and governance; and 5) contestability and redress. This study uses the UK's approach as it is representative of emerging global positioning. [2] For example, the UK's principles overlap with the EU's principles in the ethical guidelines for trustworthy AI (AIHLEG, 2019). Limitations inherent in the use of principles in regulating AI are acknowledged, for example as guidance still leaves scope for local context and judgment (Goodman et al., 2020), rather than establishing concrete measures (Coeckelbergh, 2019). We justify the use of this principles-based approach as the field is not yet mature, and therefore flexibility and context-dependence are necessary.

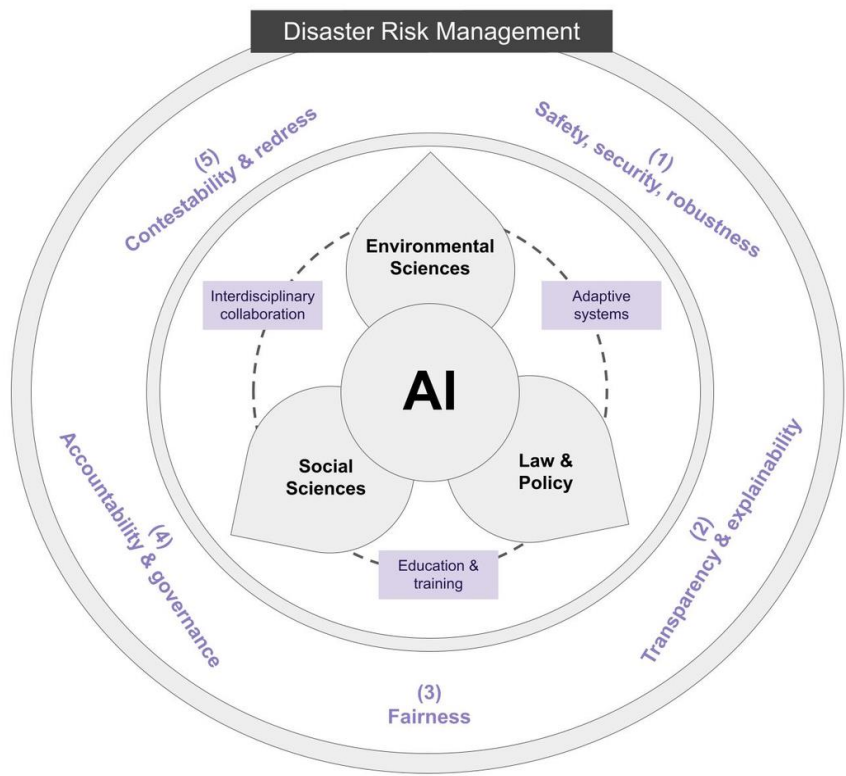

Figure 1. Five principles for regulating AI use in disaster risk management. These principles should guide multi-level actors in legal, policy-making, science and innovation sectors. Significance from these principles include: i) building adaptive data mining systems for legal and physical environments; ii) encouraging public engagement and law; and iii) forging interdisciplinary education and training in disaster management.

**Principle 1. Safety, security and robustness**

Applications of AI should function in a secure, safe and robust way where risks are carefully managed. Furthermore, the fact that AI systems can autonomously develop new capabilities can increase the risks of safety and security of their use. Use of AI and Big Data also raises significant privacy concerns. AI can process huge amounts of data that was not originally created for disaster management purposes (e.g., social media) and therefore there is a need to ensure that the data is not misused (Ufert, 2020). The EU General Data Protection Regulation defines rules on the protection of personal data but, due to rapid changes in AI technology, frequent updating of the regulation may be needed (Gunes Peschke & Peschke, 2022).

**Principle 2. Transparency and explainability**

AI systems are often initially developed as opaque 'black box' processes that are complex and difficult to understand, especially when machine learning methods are involved (Glikson & Woolley, 2020). This lack of transparency can make it challenging to determine how these Big Data and AI systems arrive at their decisions, confounding ultimate responsibility for the derived decisions (Berber & Srećković, 2023). Organisations developing and deploying AI should be able to communicate when and how it is used and explain the system's decision-making process in an appropriate level of detail that matches the risks posed by the use of AI. If AI is used to assist high-stakes decisions, such as controlling floodgates during a flood emergency, the systems need to be as transparent and explainable as possible to ensure their responsible data mining use (e.g. Papadakis et al., 2022). Nevertheless, transparency is strongly recommended to be supported by two other principles, i.e., interoperability and reusability of data. They enhance the capacity of integrating and sharing information among administrative competences alongside civil society initiatives using the available data to produce knowledge (Soylu et al., 2022; Morales et al., 2014).

**Principle 3. Fairness**

AI systems must comply with existing laws and not discriminate against individuals or create unfair outcomes. There have been some debates on the fairness of outcomes from AI-assisted processes, such as impacts on insurance offers (Lamberton et al., 2017) and recruitment outcomes (Albert, 2019). AI use in decision-making processes, particularly for high-impact outcomes, should be justifiable and not arbitrary. An AI system might unintentionally leave some communities behind, e.g., low-income, immigrant, elderly, and disabled residents, who find it difficult to access the information before and during a disaster. For example, social media data is increasingly used to provide situational awareness and support disaster response, yet may marginalise constituencies with low usage of digital media. Existing studies show uneven representation in a disaster situation and the presence of bias in social media data (Wiegmann et al., 2021).

**Principle 4. Accountability and governance**

Measures are needed to ensure there is appropriate oversight of the way AI is being used, with clear accountability for its outcomes. Key to this principle are the clear expectations for regulatory compliance on appropriate actors involved in the AI life cycle, from research and development to deployment and use. It has been known that digital infrastructures frequently fail and are prone to security issues (Lehto, 2022). For example, a flash flood in Rhineland-Palatinate, Germany, killed at least 117 people in 2021 and this was partly due to the failure of the federal government's weather warning app to notify the residents (Thieken et al., 2023; Olterman, 2021). Formulating a governance mechanism that enables the identification and mitigation of potential risks is important with a growing trend of reliance on smart systems.

**Principle 5. Contestability and redress**

The AI systems should be able to be contested by experts and stakeholders to ensure they are fair and accurate, such as in predicting the likelihood of an environmental hazard. If an AI system makes a mistake (particularly false negatives), there should be clear routes to dispute harmful outcomes or decisions generated by AI. Some ways to achieve this include having humans review the system's decision, retraining the system on new data, or implementing human review checkpoints in the decision train. Given the autonomy and opacity of AI systems, it would be difficult to understand and contest the outcomes generated or assisted by them. Linking it to the transparency and explainability principles, the overall governance mechanism needs to ensure that AI use is contestable, for example in redressing failures associated with misprediction (as in the German flood case above) or unfairness in receiving disaster relief during and following disasters (Almada, 2019; Lyons et al., 2021).

**3. IMPLICATIONS FOR DISASTER MANAGEMENT**

Drawing on insights from the co-production meetings of lawyers, social scientists, and environmental scientists and principles for responsible data mining, this section provides a summary of identified three major themes on the wider significance of the five principles presented in the previous section.

**3.1 AI for adaptive disaster management: environmental science and law**

All phases of disaster risk management must be adaptable. It is here that AI, environmental science and technology related applications, and law can combine to play a crucial role. AI and disaster management applications deal with engineering design and computational models that can be made inherently flexible. For example, an AI-generated early warning system designed by environmental scientists can issue warnings for potential disaster events. Similarly, measuring the scale of a disaster's impact in the recovery phase is also possible through remote sensing data that can be mined by AI. In the creation and utilisation of these AI and disaster management applications, the law can provide a flexible framework on how they might be used safely, transparently, and fairly to benefit the wider community while defining liability and enforcement provisions (Francesch-Huidobro, 2022; Nemakonde & van Niekerk, 2022; Villa, 2022).

However, legislation governing AI or disaster management applications -or the functions of the environmental applications themselves- must realise that these processes are dynamic, iterative and adaptive, improving their outcomes over time. For example, the design of an early warning system raises a host of issues; here we discuss three.

The first issue relates to environmental data availability and data quality for training AI. AI needs a vast amount of data to issue reliable predictions (Duan et al., 2019). However, such data are not always available in the desired quantity. To mitigate this problem, one could improve the algorithms governing the application while also clearly reporting the uncertainty of the predictions (Gneiting & Raftery, 2007). Such disclosure reveals the transparency of the system to the public, potentially improving communication and trust. This can also prompt changes to the legal framework governing AI and disaster management environmental applications. (Cutter, 2022; Kahneman & Tversky,1979). Despite advancements in the interpretability, interoperability, and reusability of AI environmental applications, there will always be some trade-off with their flexibility (James et al., 2013).

A second issue is that the outcomes of AI in environmental disaster management applications must also be equitable, as it is the poorest sectors of society who suffer most from environmental disasters (Matsuda, 2022). Assuming the quantity and quality of data are acceptable, fairness should be integrated during the AI training phase with regulations mandating and enforcing equity of use among all social strata.

Finally, early warning systems using AI raise concerns about automatic issuing of flooding predictions. AI outcomes need to be continuously controlled, but final decisions to trigger an early warning should be given by a responsible scientist or technical manager. How accountability, liability, redressability, and enforcement should be assigned remain open issues for law and social scientists to decide (Berber & Srećković, 2023).

### 3.2 Public engagement and law

In all phases of disaster management, social sciences, law, and public engagement play unique roles in regulating the use of AI and Big Data for environmental hazards. Social scientists and lawyers can help in identifying, analysing and reporting to regulators risks accompanying the use of AI and Big Data for disaster management. Examples may include the violation of a privacy rights in the use of Big Data from mobile apps, or potential discrimination between citizens regarding training, funding, implementation, and enforcement of early flood warnings and evacuation plans (Carlarne, 2022; Gable, 2022; Matsuda, 2022; Sherwin, 2022). There is the need to ensure that AI-generated policies for promoting resilience to disasters, which do not conflict with prevailing local laws, geography, and historic forms of development (Marshall, 2015; Finn & Marshall, 2018). Social scientists, legal regulators, and the public must also work together to formulate and communicate the ethical principles (e.g. fairness, transparency, safety, accountability, and contestability) that should guide AI use for adaptive disaster management.

Public engagement is crucial for the efficacy of environmental disaster management systems. It is consequently critical that disaster management agencies and decision-makers liaise with the public frequently and clearly as these people are on the frontlines of disaster, and individual citizens often also can identify timely, credible solutions to address immediate risks (Peliño-Golle & Baula, 2022).

Social media provides multiple platforms through which the public can constantly interface with legal regulators (see Figure 2 for the social media landscape and emerging AI and Big data considerations in disaster management). 'Crowdsourcing' has become a common way for interested members of the public to respond to informational needs of organisations or agencies (Rowberry, 2022). Such public-to-government exchange does, however, pose challenges. One issue is determining whether the use of such public data is truly representative as the dataset would include only information from the digitally able (Lieske et al., 2019). Another is that such exchange could lead to the collection of substantial volumes of data (Ghani et al., 2019), which may impact privacy and safety concerns of the public. Nevertheless, several nations and government entities already collect disaster data and are grappling with these legal issues. In India, the Disaster Management Act 2005 (Government of India, 2005), provides a comprehensive disaster management legal framework that is compared with similar legal frameworks (e.g. Madan & Routray, 2015; Mohanan & Menon, 2016; Shakeri et al., 2021). India's law states that the government may collect and share personal information in the event of a disaster, but only if there is a 'reasonable' justification for doing so. The much-discussed General Data Protection Regulation (GDPR), promulgated by the European Union (see discussions by Tikkinen-Piri et al., 2018; Hoofnagle et al., 2019) is another example of how laws might regulate the processing of personal data from social media platforms for use in disaster management.

As more countries begin to use AI in adaptive disaster management, involving the public in honest, robust, representative and constant discourse will be crucial to engendering trust and, ultimately, to saving lives and reducing property damage (Cutter, 2022).

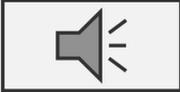

Figure 2. The potential of social media and big data to enhance disaster risk management, highlighting key considerations for effective implementation

## 3.3 Interdisciplinary education and training

Disaster management for environmental hazards involves substantial technical work of a diverse nature, such as data collection and analysis, hazard modelling and crisis management. However, disaster management cannot be successful without effective laws and policies at national, regional or international scales (Seneviratne et al., 2010). Interdisciplinary studies are becoming more common and urgent to resolve long-standing and emerging environmental issues, particularly issues such as disaster management that span multiple disciplines. Consequently, engendering collaboration and co-learning between environmental and social scientists in conjunction with legal and policy experts is essential. One pioneering interdisciplinary collaboration is *The Cambridge Handbook of Disaster Law and Policy* (Kuo et al., 2022), which provides a foundational resource for exploring various legal and policy frameworks for managing environmental disasters, including case studies from across the globe.

International and interdisciplinary cooperation and coordination are essential for effectively managing future disasters. This is due to both the global nature of climate justice but also contributory human influence on many contemporary disasters. Furthermore, disasters have no political boundaries, with potential cascading impacts across regional, continental and global scales (Cutter, 2018; Gill, 2022), including 'spillover' effects from measures in one country upon others, such as unconstrained carbon emissions, forest felling, pollutant emissions, fish stock overexploitation. Authoritative voices, such as UNESCO, the League of European Research Universities (LERU) and the British Academy, are increasingly emphasizing the importance of interdisciplinary research and teaching to unleash the potential of universities to achieve interdisciplinarity in practice.

In an educational context, increasing use of AI can play an important role by bringing together students from different disciplines to devise interdisciplinary solutions to disaster management problems. This can help to ensure that future professionals develop a broad understanding of the issues involved in disaster management and are better equipped to develop creative and innovative interdisciplinary solutions. Disaster law and policy should be included in the curricula of university degrees on environmental sciences and law (Baker et al., 2022). Georgia State University provides an important example of this integrated approach by offering an interdisciplinary "Urban Environmental Sustainability" program that links lawyers, environmental scientists, economists, and disaster management specialists to tackle multi-faceted climate issues facing municipalities worldwide.

## 4. CONCLUSION

The interdisciplinary and collaborative framing of this study reflects on the why's and how's of establishing a system of practice to support and accelerate adaptive disaster management through AI and data mining, emphasising how law and policy can help regulate its responsible and ethical use. The twenty-first century has witnessed a speedy evolution of AI technologies, coincident with increasing global intensity and frequency of environmental hazards. AI for disaster management has achieved a certain level of technical maturity, highlighting the need for greater attention on its use addressing issues such as bias, transparency, and privacy, which pose significant ethical and legal risks for all disaster management stakeholders at different levels of competencies and responsibilities. Although foundational precepts to inform the further evolution of disaster law and policy have been achieved, significantly including the five principles discussed here, there is still a question of how these principles can be translated into enforceable regulations. Governments are therefore strongly urged to develop regulatory frameworks to ensure that the benefits of AI can be harnessed whilst minimizing unintended negative outcomes, necessitating interdisciplinary efforts to consider the needs and rights of all stakeholders impacted by environmental hazards.

The use of AI in disaster management is a complex issue involving multiple factors related to the different views of environmental scientists and decision-makers about the usefulness and accuracy of AI for representing natural and human systems. Whilst collaboration among professionals from relevant disciplines and communities who are affected, directly or indirectly, should be promoted, these diverse views are currently hindering the implementation of AI in disaster management frameworks. Bridging these conceptual differences may enable improved adaptation of AI technologies to reflect local contexts, laws, and needs for natural resources considering the wider ramifications across a broad spectrum of human activities, including making progress towards Sustainable Development Goals (SDGs) (Vinuesa et al., 2020; Costanza et al., 2023).

Ethics consideration should be part of AI development and should be undertaken appropriately, not performatively. Creating a system that allows ethical issues to be assessed proportionately will avoid seeing this requirement as a barrier in development, but instead as a mechanism ensuring that risks have been identified and addressed as much as possible.

The use of social networks for mitigating environmental risks based on AI raises diverse privacy and liability concerns, requiring careful consideration before their use for this purpose. AI also has an important role to play in education, bringing together students from different disciplines to support the next generation of disaster managers. They will be prepared to challenge unfair, obscure, and unsafe AI decisions and equipped to develop creative and innovative solutions.

The use of AI and Big Data can enhance disaster management, but ethical considerations and safeguards are crucial to address concerns around bias, transparency, accountability and privacy. Outreach and education are also essential in building resilient communities now and training communities to be resilient in the future. We propose to integrate an adaptive legal framework based on state indicators derived from environmental Big Data networks to inform nature-based solutions for regional planning and sustainable management decisions. Overall, these adaptive data mining approaches based on the intersectional priorities of legal and environmental systems can help reduce the impact of disasters and create more sustainable and resilient communities.

**Funding information:** KC and TO work together on the Royal Society project on "Spatiotemporal Variation Characteristics of Compound Dry and Hot Events and Their Impacts on Vegetation Growth Across the Mid-latitudes of Eurasia". KC is supported by the Vice Chancellor's Accelerator Programme Award (2022-2024) to develop AI and Big Data approaches for extracting climate and weather information from convection-permitting models for environmental management in urban and green and blue spaces. He is also an awardee, along with MD and TF, for the Vice Chancellor's Challenge Fund (2023-2024) "VIS-Studio: An Immersive Reality and AI Solution for Data Visualization to Support Collaborative Decision-Making for Extreme Weather and Disaster Scenarios". TO is Vice Chancellor's Earlier Career awardee for Responsible AI.

**Conflict of interest:** The authors confirm that there is no conflict of interest to declare.

**Footnote**

1. The co-production process underpinning this paper was initiated in law conferences bringing together a lawyer network, leading to publication of *The Cambridge Handbook of Disaster Law and Policy* in summer 2022: a valuable resource for understanding international legal and policy frameworks for managing environmental disasters. This study used a variety of techniques to involve different communities, including filling tables (see Supplementary materials), sharing documents, and conducting focus groups. The evolving work has been presented at writing retreats hosted by the Centre for Water, Communities and Resilience (CWCR) and the School of Architecture and Environment by the University of the West of England. In the spring of 2023, we presented our work to the South American Action Group communities with the National Center for Atmospheric Research (NCAR) researchers and the Global Institute of Water Security researchers from the University of Saskatchewan. We have also tested our work in our Water and Energy Future classes for the Geography and Environmental Management programmes during the 2022-2023 academic year, designed to engage the next generation of disaster managers. The co-production processes entailed in development of this paper emphasise the need for an interdisciplinary approach to disaster management.

2. We use the UK's principles here as they arguably represent general positions of other countries and organisations. However, we also acknowledge that some countries and organisations may put different emphasis on their initiatives. For example, the US emphasises the use and development of responsible AI for *their citizens.*